\documentclass[prd,aps,twocolumn,nofootinbib,preprintnumbers,superscriptaddress,preprintnumbers,balancelastpage,longbibliography]{revtex4-2}
\usepackage[english]{babel}

\usepackage{amsmath,mathtools,physics,xfrac}
\usepackage[normalem]{ulem}
\usepackage{graphicx}
\usepackage{afterpage}
\usepackage{float}
\usepackage{subfigure}
\usepackage{rotating}
\usepackage{multirow}
\usepackage{tabularx}
\usepackage{booktabs}
\usepackage{fancyhdr}
\usepackage[utf8]{inputenc}
\usepackage{theorem}
\usepackage{moreverb}
\usepackage{euscript}
\usepackage{psfrag}
\usepackage{slashed}
\usepackage{mathtools}
\usepackage{makecell}
\usepackage{adjustbox}
\usepackage{dcolumn}
\usepackage{bm}
\usepackage[dvipsnames]{xcolor}
\usepackage{graphics}
\usepackage{hyperref}

\hypersetup{
     colorlinks   = true,
     citecolor    = Cyan,
     urlcolor     = Cyan,
     linkcolor    = Cyan
}

\usepackage{color,xcolor}

\definecolor{darkblue}{rgb}{0.0,0.0,0.75}
\definecolor{darkred}{rgb}{0.6,0.0,0}
\definecolor{darkgreen}{rgb}{0.0,0.6,0.}

\newcommand{\K}{\,\mathrm{K}}
\newcommand{\W}{\,\mathrm{W}}

\newcommand{\fW}{\,\mathrm{fW}}

\newcommand{\mum}{\,\mu\mathrm{m}}

\newcommand{\Wmum}{\,\mathrm{W}\mu\mathrm{m}^{-3}}

\newcommand{\eV}{\,\mathrm{eV}}
\newcommand{\GeV}{\,\mathrm{GeV}}
\newcommand{\MeV}{\,\mathrm{MeV}}
\newcommand{\keV}{\,\mathrm{keV}}
\newcommand{\meV}{\,\mathrm{meV}}
\newcommand{\mueV}{\,\mathrm{\mu eV}}

\newcommand{\cmsq}{\,\mathrm{cm^{2}}}

\newcommand{\mchi}{m_\chi}
\newcommand{\nchi}{n_\chi}
\newcommand{\bq}{\boldsymbol{q}}
\newcommand{\oq}{\omega_{\boldsymbol{q}}}
\newcommand{\bv}{\boldsymbol{v}}
\newcommand{\fmed}{F_\mathrm{med}}
\newcommand{\sigman}{\sigma_{\chi N}}
\newcommand{\qbz}{q_\mathrm{BZ}}
\newcommand{\nqp}{n_\mathrm{qp}}

\newcommand\redsout{\bgroup\markoverwith{\textcolor{red}{\rule[0.5ex]{2pt}{0.4pt}}}\ULon}

\begin{document}

\preprint{SLAC-PUB-17691}

\title{Dark Matter Induced Power in Quantum Devices}

\author{Anirban~Das}
\thanks{{\scriptsize Email}:
\href{mailto:anirband@slac.stanford.edu}{anirband@slac.stanford.edu}
}
\affiliation{SLAC National Accelerator Laboratory, 2575 Sand Hill Rd, Menlo Park, CA 94025, USA}

\author{Noah~Kurinsky}
\thanks{{\scriptsize Email}: \href{mailto:kurinsky@slac.stanford.edu}{kurinsky@slac.stanford.edu}
}
\affiliation{SLAC National Accelerator Laboratory, 2575 Sand Hill Rd, Menlo Park, CA 94025, USA}
\affiliation{Kavli Institute for Particle Astrophysics and Cosmology, Stanford University, Stanford, CA 94035, USA}

\author{Rebecca~K.~Leane}
\thanks{{\scriptsize Email}: \href{mailto:rleane@slac.stanford.edu}{rleane@slac.stanford.edu}
}
\affiliation{SLAC National Accelerator Laboratory, 2575 Sand Hill Rd, Menlo Park, CA 94025, USA}
\affiliation{Kavli Institute for Particle Astrophysics and Cosmology, Stanford University, Stanford, CA 94035, USA}

\date{March 23, 2024}

\begin{abstract}
We point out that power measurements of single quasiparticle devices open a new avenue to detect dark matter (DM). The threshold of these devices is set by the Cooper pair binding energy, and is therefore so low that they can detect DM as light as about an MeV incoming from the Galactic halo, as well as the low-velocity thermalized DM component potentially present in the Earth. Using existing power measurements with these new devices, as well as power measurements with SuperCDMS-CPD, we set new constraints on the spin-independent DM scattering cross section for DM masses from about 10 MeV to 10 GeV. We outline future directions to improve sensitivity to both halo DM and a thermalized DM population in the Earth using power deposition in quantum devices.
\end{abstract}

\maketitle

\noindent\textit{\textbf{Introduction.---}} At any given moment, a powerful stream of DM particles from the Galactic halo flows into the Earth. This Galactic DM has been extensively searched for in direct detection experiments, which aim to detect recoil events when DM scatters off the Standard Model (SM) target material, thereby providing a test of the DM-SM scattering cross section. Typically, the energy threshold of direct detection experiments assuming nuclear recoils is about a keV, corresponding to the recoil expected for DM with mass above about a GeV for standard analyses~\cite{LUX-ZEPLIN:2022qhg}, or MeV-scale masses when exploiting the Migdal effect~\cite{Vergados:2005dpd, Moustakidis:2005gx, Bernabei:2007jz, Ibe:2017yqa,XENON:2019zpr} or electron recoils~\cite{Tiffenberg:2017aac,SENSEI:2020dpa,XENON:2019gfn}.

Given the lack of a conclusive DM detection with direct detection experiments so far, interest in novel detection strategies and new devices has exploded in the last few years~\cite{Hochberg:2022apz}. In particular, the race down to increasingly low thresholds has inspired use of new detectors, including superconductors~\cite{Hochberg:2015pha, Hochberg:2015fth, Hochberg:2021pkt, Hochberg:2021ymx,Hochberg:2019cyy,Chiles:2021gxk}, superfluids~\cite{Schutz:2016tid, Knapen:2016cue, Caputo:2019cyg}, polar crystals~\cite{Griffin:2018bjn, Knapen:2017ekk, Cox:2019cod}, topological materials~\cite{Sanchez-Martinez:2019bac}, and Dirac materials~\cite{Hochberg:2017wce, Geilhufe:2018gry, Geilhufe:2019ndy, Coskuner:2019odd}. Superconductors show exceptional promise, due to their superconducting energy gaps as low as about an meV, allowing probes of light DM.

The goal of lower threshold experiments to date has been to push down sensitivity to lower DM masses, and we will exploit this to test incoming halo DM down to the MeV-scale. Lowered thresholds open up a new probe of a DM component other than the usually-considered halo DM. When the Galactic halo DM enters the Earth, it scatters, loses energy, and can become gravitationally captured. Over time, this builds up a thermalized population of DM particles bound to the Earth. For DM around a few GeV that is in local thermal equilibrium, the density of bound DM at Earth's surface can in fact be enormous: about 15 orders of magnitude higher than the local DM halo density~\cite{Neufeld:2018slx,Pospelov:2020ktu, Pospelov:2019vuf, Rajendran:2020tmw,Xu:2021lmg,Budker:2021quh, McKeen:2022poo,Billard:2022cqd, Leane:2022hkk}. Unfortunately this large density enhancement is lost on traditional direct detection experiments, as the bound DM population has a very low velocity compared to halo DM, requiring thresholds of less than about $0.05~$eV at Earth's surface.

\begin{figure}[!t]
\centering
\includegraphics[width=0.8\columnwidth]{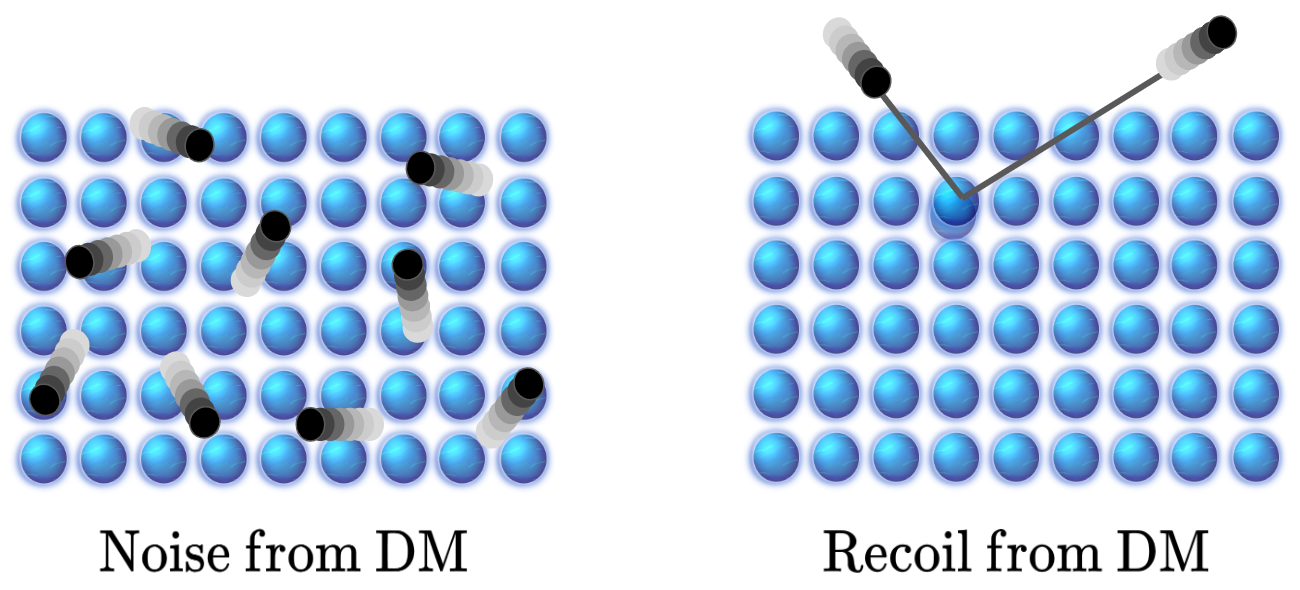}
\caption{The qualitative difference between our proposal and a conventional DM direct detection experiment. The noise arises from \textit{frequent} interaction between DM and the nuclei in the detector, as opposed to \textit{once-in-a-while recoil} of a nucleus from DM scattering.}
\label{fig:cartoon}
\end{figure}

We will demonstrate for the first time that power measurements using new quantum devices can be used to detect DM with low energy depositions. This includes sensitivity to both light DM from the halo, as well as thermalized bound DM. As schematically shown in Fig.~\ref{fig:cartoon}, for thermalized DM our proposal exploits their high DM density and is sufficiently sensitive despite low thermal velocities, compared to traditional direct detection, which only measures the less frequent and higher-velocity DM halo interactions. We point out and will use the fact that both halo DM and thermalized DM would produce excess quasiparticle generation in single quasiparticle devices, and excess power produced in athermal phonon sensors, to set new constraints on DM with interaction cross sections larger than about $10^{-34}-10^{-28}$~cm$^2$ for DM masses of $\sim300$ MeV$-10$ GeV for thermalized DM. For halo DM, we will set constraints down to about $10^{-29}-10^{-26}$~cm$^2$ for DM masses of $\sim10$ MeV$-10$ GeV.\\

\noindent\textit{\textbf{Dark Matter at Earth's Surface.---}} At Earth's position, there are two potential DM components present, which have different DM velocity and density assumptions. We will test both of these components. One is DM incoming from the Galactic halo, which is usually assumed for direct detection experiments. The other is the thermalized DM component. This thermalized component exists as once DM enters the Earth, it can thermalize, and become captured and bound to the Earth. For sufficiently large DM-SM scattering cross sections (larger than about $10^{-35}$~cm$^2$), the DM rapidly thermalizes and is said to be in local thermal equilibrium with the surrounding SM matter. In this case, the DM radial profile within the Earth, $n_\chi$, is dominantly governed by the differential equation~\cite{Leane:2022hkk}
\begin{align}
    \label{eq:radialDM}
    \frac{\nabla n_\chi}{n_\chi} + \left(\kappa+1\right) \frac{\nabla T}{T} +  \frac{m_\chi g}{T}=\frac{\Phi}{n_\chi D_{\chi N}}\frac{R_\oplus^2}{r^2}\, ,
\end{align}
where $T$ is the Earth's radial temperature profile at position $r$, $R_\oplus$ is Earth's radius, $m_\chi$ is the DM mass, $g$ is gravitational acceleration, and $\Phi$ is the incoming flux of DM particles from the Galactic halo. $D_{\chi N}\sim \lambda v_{\rm th}$ and $\kappa\sim-1/[2(1+m_{\chi}/m_{\rm SM})^{3/2}]$ are diffusion coefficients~\cite{Leane:2022hkk}, with $\lambda$ the DM mean free path, $v_{\rm th}$ the DM thermal velocity, and $m_{\rm SM}$ the SM target mass. The DM density profile is normalized by enforcing that its volume integral equals the total number of particles expected within the Earth~\cite{Leane:2022hkk}.

Solving Eq.~(\ref{eq:radialDM}) for $n_\chi(r)$ reveals that this thermalized population of DM can be significantly more abundant at the Earth's surface than the incoming halo DM particles. For DM masses around a GeV, the local DM density can be as high as about $\sim10^{14}~$cm$^{-3}$. However, as this population is thermalized within the Earth, its velocity is low. We approximate the thermalized DM velocity distribution as a truncated Maxwell-Boltzmann distribution,
\begin{align}
    f_\chi(\bv) = \frac{1}{N_0} e^{-(\bv/\bv_{\rm th})^2}
    \Theta(v_\mathrm{esc}-v)\,,
\label{eq:velocity}
\end{align}
where $N_0$ normalizes the distribution, and $v_{\rm th}^2=8T_\chi/\pi\mchi$ with $T_\chi \simeq 300\K$. This velocity would require thresholds of $E\lesssim0.05~$eV for conventional detection techniques. This is much lower than the reach of typical direct detection experiments, and so requires new techniques to be detected. Our assumption of DM being at room temperature of $\sim300$ K is reasonable, as even at the largest cross sections considered the mean free path is much larger than the size of our devices, such that DM is not expected to thermalize with the device itself.

For halo DM, in Eq.~(\ref{eq:velocity}) $v_\mathrm{th}$ is replaced by the average DM velocity in the halo $v_0=230$~km/s. In this case, the relative velocity between the Earth and DM also becomes important. Hence, for halo DM we use the boosted velocity $\bv \to \bv+\bv_\oplus$ in Eq.~(\ref{eq:velocity}), where $|\bv_\oplus|=240\,\mathrm{km/s}$ is the Earth's velocity in the galactic rest frame. The halo DM density is assumed to be $0.4\,\mathrm{GeV\,cm^{-3}}$. We now show that quantum devices are highly sensitive to DM with low energy depositions through their power measurements, which includes both the thermalized DM population, as well as light halo DM.\\

\noindent\textit{\textbf{Scattering Rate \& Energy Deposition.---}} As a DM particle with velocity $\bv$ scatters in the detector and transfers momentum $\bq$, it deposits an amount of energy 
\begin{align}\label{eq:DM_phase_space}
    \oq = \bq\cdot\bv - \frac{q^2}{2\mchi} = E_f-E_i\,.
\end{align}
As a result, the target makes a transition from $\ket{i}$ to $\ket{f}$. For such low energy depositions, the momentum transferred is comparable to the inverse size of nuclear wavefunction in a detector crystal, and the inter-atomic forces become important. Hence, lattice vibrations or phonon excitations will be used to compute the DM scattering rate. The total rate per unit target mass can be written as\,\cite{Kahn:2021ttr,Trickle:2019nya}
\begin{align}\label{eq:rate}
    \Gamma = \frac{\pi\sigman\nchi}{\rho_T\mu^2} \int d^3v f_\chi(\bv) \int \frac{d^3q}{(2\pi)^3} \fmed^2(q) S(\bq,\oq)
\end{align}
Here, $f_\chi(\bv)$ is DM velocity distribution, $\rho_T$ is the target density, $\sigman$ is the DM-nucleon scattering cross section, $\mu$ is the reduced mass of the DM-nucleon system, $\fmed(q)$ is a form-factor that depends on the mediator (we assume $\fmed(q)=1$), and $S(\bq,\oq)$ is the dynamic structure factor containing the detector response to DM scattering and depends on the crystal structure of the target material. 

To compute DM scattering rates, we follow Refs.~\cite{Knapen:2021bwg,Campbell-Deem:2022fqm} and use the publicly available code DarkELF. We modify DarkELF in two main ways. Firstly, we update the local DM density and DM velocity input to be that described in the previous section, for halo or thermalized DM as appropriate. Secondly, the code was developed only for materials with two atoms per primitive cell, which is the smallest crystal unit. Thus, we adapt it for materials like Al which has only one atom in its primitive cell.\\

\noindent\textit{\textbf{Detection Mechanisms and Materials.---}} Detecting light halo DM or the captured DM population of low thermal energy demands use of low threshold quantum sensors that can detect $\sim \mathcal{O}(10)\meV$ energy deposition. Such sensors are usually designed using superconducting materials, which have small energy gaps\,\cite{2013NatCo...4.1913R,QCD,Fink_2021,Ren_2021}. Aluminum (Al) is a widely used superconductor for such a purpose and its characterization data is readily available. Such a small amount of energy transfer is not sufficient for nuclear recoil or electronic ionization, however DM can excite collective modes, such as phonons in the material, resulting in an excess power. For example, in one experimental setup, a bias circuit stabilizes the absorber material at its transition temperature $T_c$, where its resistance is very sensitive to any energy deposition. The total power deposited in the detector by DM in the form of phonons is
\begin{align}\label{eq:P_DM}
    P_\mathrm{DM} = \epsilon\int d\omega~\omega\dfrac{d\Gamma}{d\omega}\,,
\end{align}
where $\epsilon$ is an efficiency factor that depends on the experimental setup. We will use this to calculate excess power due to DM and set constraints on DM-SM interactions. 
Volume-scaled detectors based on conventional semiconductors, such as Si, can also be used as the absorber material to look for ambient power deposition; the power deposited per unit volume can be obtained from Eq.~(\ref{eq:P_DM}).

We also consider excess quasiparticle production from DM. In a superconducting metal, the electrons are bound into Cooper pairs through a long-range interaction with phonons. When a DM particle scatters with a nucleus, it may deposit its kinetic energy in the form of phonons. If the deposited energy exceeds the energy gap $\Delta$ of the superconductor, these excess phonons will break some of the Cooper pairs and release quasiparticles above the gap. We will therefore set limits on DM-SM interactions by calculating quasiparticle production rates from DM.

The quasiparticle generation rate $R_{\rm qp}$ by DM scattering can be written as
\begin{align}\label{eq:qp_gen}
    \nonumber R_{\rm qp} &= \frac{\epsilon_{\rm qp}}{\Delta}\int d\omega~ \omega\,\frac{d\Gamma}{d\omega} \\
    &\approx \left(\frac{P_{\rm DM}}{9\times 10^{-23} \Wmum}\right) \,\mathrm{Hz\,\mu m^{-3}}\,,
\end{align}
where $P_{\rm DM}$ is the deposited DM power above the gap in $\Wmum$, assuming a 60\% quasiparticle generation efficiency ($\epsilon_{\rm qp}=0.6$)~\cite{Kaplan,Hochberg:2021ymx}, and using $\Delta \simeq 340\mueV$ for Al.

A conservative estimate of $\nqp$, the steady-state quasiparticle density, can be found using mean field results from Ref.~\cite{bespalov} as follows,
\begin{equation}
    \dfrac{d\nqp}{dt} = -\Gamma_R - \Gamma_T + A \approx -\Bar{\Gamma}\nqp^2 - \Bar{\Gamma}_T\nqp + A\,,
\end{equation}
with $\Gamma_R, \Gamma_T, A$ as the recombination, trapping, and generation rates, respectively. With a steady state injected power density $P$, we have $A=P/(2\Delta)$ where $\Delta$ is the gap energy. In equilibrium, we thus find
\begin{equation}
P/(2\Delta) = \Bar{\Gamma}\nqp^2 + \Bar{\Gamma}_T\nqp \,.
\end{equation}
The mean field calculation assumes no trapping of QPs with $\Bar{\Gamma}_T=0$, which leads to the relation $\nqp=\sqrt{A/\Bar{\Gamma}}\propto \sqrt{P}$.
In case of DM scattering, $A=R_{\rm qp}$ using Eq.~(\ref{eq:qp_gen}), and $\bar{\Gamma}=40\,\mathrm{s^{-1}\mu m^3}$ for Al. The steady-state density is therefore
\begin{equation}\label{eq:qp_density}
    \nqp \approx \left(\frac{P_{\rm DM}}{3.6\times 10^{-21}\mathrm{W}}\right)^{1/2}\mum^{-3},
\end{equation}
which can be compared to known measurements to set new constraints. We now discuss devices that can be used to detect DM using power deposition.\\

\noindent\textit{\textbf{Detecting Dark Matter with Single Quasiparticle Devices.---}} 

 \textit{(i) Quasiparticle Tunneling in Transmon Qubits:} Quasiparticle excitations formed from broken Cooper pairs are important to minimize in quantum devices, as the quasiparticle background limits the operation of applications such as radiation detectors and superconducting qubits. To study the effect of quasiparticle tunneling on the decoherence of a transmon qubit, Ref.~\cite{2013NatCo...4.1913R} constructed a single junction superconducting qubit made of Al, and studied its decoherence by monitoring single-charge tunneling rate. From the observed relaxation rate of the qubit, they found a quasiparticle density of $0.04\pm0.01~\mum^{-3}$ with a thermalized distribution~\cite{2013NatCo...4.1913R}.

 We convert this measurement to a power density using Eq.~(\ref{eq:qp_density}), finding an upper limit of $3.92\times 10^{-24}\Wmum$.
 We compare this directly with the expected DM induced quasiparticle density in Al, and consider this an upper limit on residual power injection. Moreover, the source of this quasiparticle density is not known and usually assigned to the background radiation from the environment\,\cite{2013NatCo...4.1913R}. Therefore, it is possible that the DM scattering contributes to it too. We overall point out that quasiparticles produced by DM, and therefore the DM-SM scattering rate, can be probed using devices with low-quasiparticle density backgrounds.  

\textit{(ii) Low Noise Bolometers:} Understanding our Universe deep into the infrared would reveal new secrets of galaxy formation, exoplanets, and so much more. However, far-infrared spectroscopy requires new cryogenic space telescopes with technologies capable of measuring very cold objects, and therefore require low noise equivalent power in their detectors. Adapting technology from quantum computing applications, Ref.~\cite{QCD} developed a quantum capacitance detector where photon-produced free electrons in a superconductor tunnel into a small capacitive island. This setup is embedded in a resonant circuit, and therefore can be referred to as a ``quantum resonator". This quantum resonator measured excess power of $4\times 10^{-20}\W$~\cite{QCD}, making it the most sensitive existing far-infrared detector. The volume of the absorber used in this case was a mesh grid, roughly 60 microns square with a 1\% fill factor and 60~nm thick. This corresponds to a volume of around $1.56\mum^3$ and thus a power density measurement of $2.6\times 10^{-20}\Wmum$.  We therefore point out that single quasiparticle devices can be used for DM detection through their power measurements, and will use this current measurement to set constraints on the DM-SM scattering rate which would produce excess power. Note that this detector has a calibrated photon detection efficiency of greater than $95\%$, which reduces the systematic uncertainty on the power limit. Ref.~\cite{QCD} excludes the possibility that this is induced by residual radiation and represents a true, measured excess power.

For both the devices above, we will only consider DM detection from the superconductor films, rather than the substrate. In case of the transmon qubit, it is a conservative choice. While for the bolometer, the substrate is connected to the ground plane made of metallic Au-Ti, which acts as a phonon trap stopping the phonons created in the substrate from travelling into Al; see the Supplementary Material for more details.

\noindent\textit{\textbf{Detecting DM Power Deposition with Existing DM Detectors.---}} While we have pointed out new devices that can be used as DM detectors above, we also point out that more conventional quantum sensors with volume-scaling can already be used to constrain low-energy deposition DM through their power measurements. We consider SuperCDMS detectors, which have recorded volume-scaled transition-edge sensor (TES) bias power measurements in which the TES is coupled to a large aluminum absorber \cite{Ren_2021}. For Ref.~\cite{Ren_2021} we find a bias power of $2\times 10^{-15}\fW$, and an Al absorber with a volume of $2\times 10^{6} \mum^3$. This yields a power density of order $10^{-21}\Wmum$. The coupling efficiency of power to the readout in this case is 30\% ($\epsilon=0.3$), so our bound on DM power using Al would be $3\times 10^{-21}\Wmum$. However, the best constraints on DM scattering power injection come from SuperCDMS-CPD~\cite{Fink_2021}, which instead has 10.6\,g of Si as the absorber material. In this case, an excess power of 6\,pW was measured in the phonon sensor arrays, corresponding to an excess substrate power of 18\,pW or $10^{-24}\Wmum$. As this provides the superior limit, we use the measurement from SuperCDMS-CPD~\cite{Fink_2021}. Using the Si as the sole absorber volume in the phonon calorimeter device used in SuperCDMS is justified, because its design was optimized to maximize the collection of phonons from the Si block into the superconducting Al fins; see the Supplementary Material for further discussion. Note that in Ref.~\cite{Billard:2022cqd}, future projections with a hypothetically altered version of SuperCDMS were considered for thermalized DM. Here, we already use the current SuperCDMS measurements to set the first limits.\\

\noindent\textit{\textbf{New Dark Matter Constraints.---}} Figure~\ref{fig:limits} shows the new bounds we derive for spin-independent DM-SM scattering. The strongest sensitivity is achieved using quasiparticle density measurements. The conversion from quasiparticle density to quasiparticle generation rate can only be trusted to an order of magnitude, and so we show
two orange “quasiparticle” lines, representing a conservative and an optimistic constraint, which corresponds to taking $\overline{\Gamma} = 4$ or $400\,\mathrm{s^-1}\mum^3$ respectively, i.e. moving the quasiparticle generation rate between its expected range of validity. The next strongest bound arises from scattering power injection with SuperCDMS CPD, shown in magenta, where we find that their volume advantage still overcomes the superior power sensitivity of the low-noise bolometer, which is too weak to show on the plot for halo DM. The top two bounds correspond to limits using the incoming halo DM, while the blue “thermalized DM” constraint uses only the thermalized DM population.

\begin{figure}[t!]
    \centering
    \includegraphics[width=\columnwidth]{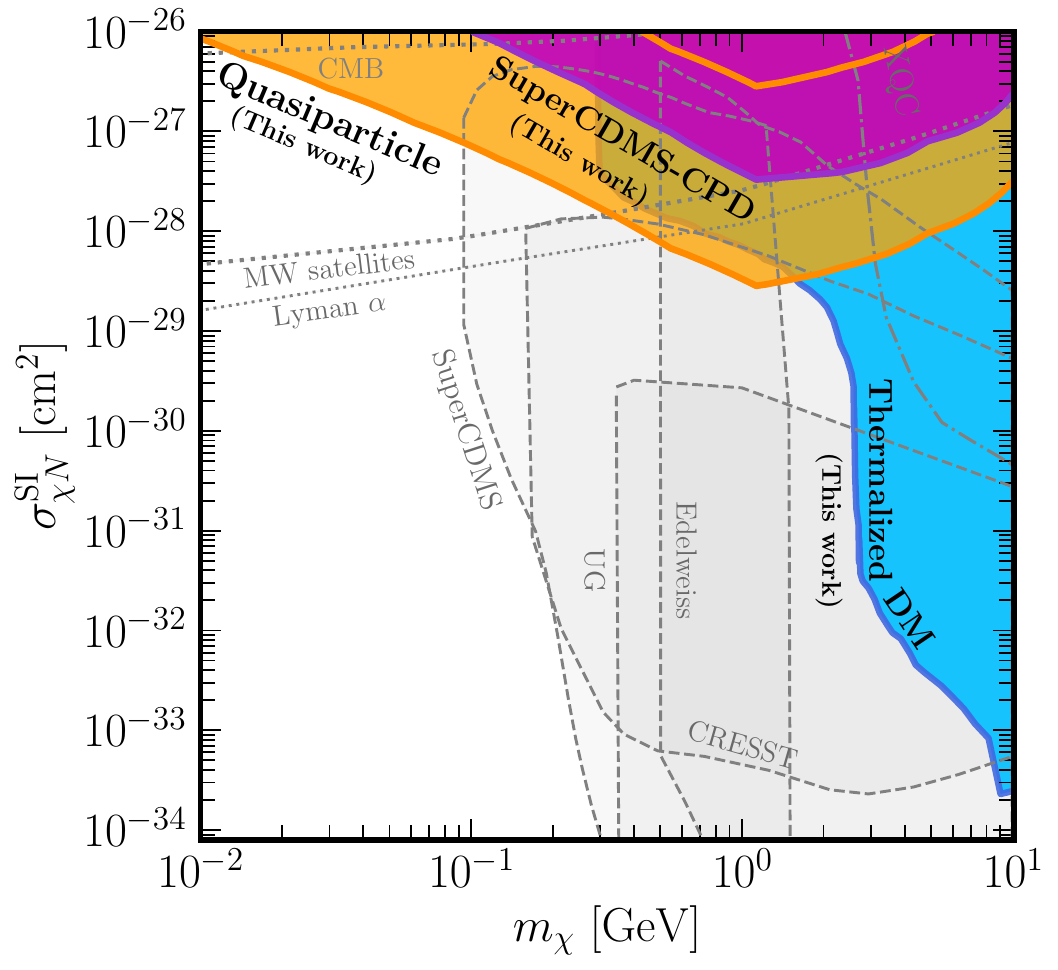}
    \caption{New limits on spin-independent DM-nucleon scattering cross section $\sigman$ derived in this work, using quantum devices. We show halo DM limits from quasiparticle production measurement (orange) and SuperCDMS-CPD (magenta). ``Thermalized DM" (blue) is our new constraint on the thermalized DM population, with several experiments overlapping in their exclusion of this population. The gray regions are other existing limits, see text for discussion.}
    \label{fig:limits}
\end{figure}

In Fig.~\ref{fig:limits}, for thermalized DM, all three of our quantum devices overlap in their constraint strength. 
The thermalized DM curve is truncated not due to device sensitivity, but rather because DM evaporates to the left of the blue contour. This occurs due to thermal kicks transferring too much energy to the DM particle relative to the gravitational binding energy of the Earth, such that DM escapes the Earth and does not remain bound to produce any signal. If there was no DM evaporation, as is possible in models other than the purely contact cross section setup we considered here~\cite{Acevedo:2023owd}, the quantum sensors would have sensitivity extending to much lower cross sections for the thermalized DM component. This motivates further studies of models that do not evaporate at these DM masses and cross sections. In Fig.~\ref{fig:limits}, for thermalized DM, largest DM densities are achieved for the asymmetric DM, which we assume for our densities, though we are also sensitive to annihilating DM models; $p$-wave annihilating DM does also not affect our assumed DM densities~\cite{Leane:2022hkk}. Note that in comparison, the incoming halo DM limits do not require any assumptions about DM annihilation. The halo limits extend to lower masses as incoming DM which later evaporates still leads to a bound from when DM is first entering the Earth. We do not show ceilings for our limits, which will exist but require simulations to calculate accurately. We expect this could lower the limits at large cross sections at the top of Fig.~\ref{fig:limits}. We focus on below 10 GeV in Fig.~\ref{fig:limits}, as heavier DM more readily sinks further into the Earth, and so the sensitivity becomes weaker outside this mass range.

In Fig.~\ref{fig:limits} we also compare with existing limits, including those from astrophysical systems such as Milky Way satellites~\cite{DES:2020fxi}, Lyman-alpha~\cite{Rogers:2021byl}, and the CMB~\cite{Xu:2018efh}. There are also lab experiments overlapping with part of our parameter space, namely CRESST~\cite{CRESST:2017ues}, SuperCDMS~\cite{SuperCDMS:2020aus}, Edelweiss~\cite{EDELWEISS:2019vjv}, XQC~\cite{Mahdawi:2018euy}, and ``UG" which is a combined limit line from deep underground experiments~\cite{CRESST:2017cdd, SuperCDMS:2017nns, LUX:2016ggv, XENON:2017vdw, PandaX-II:2017hlx, Hooper:2018bfw}. However, there is significant ambiguity in the interpretation at cross sections exceeding about $10^{-30}$~cm$^2$ where the Born approximation breaks down, and the nuclear coherence across different detector materials is not well defined without using a DM model~\cite{Digman:2019wdm,Xu:2020qjk}. For transparency we show all bounds that have been quoted in this parameter space, but emphasize many of these bounds are not generic, have different assumptions, and cannot be directly compared in a consistent manner without a DM model~\cite{Digman:2019wdm,Xu:2020qjk}. As such, our bounds significantly add to the picture of exclusions on this parameter space, even in the regions where they naively appear to overlap. There are also regions where we only overlap with astrophysical measurements, which are inherently less certain than our lab-based measurements. In addition, these astrophysical bounds disappear for models where the DM tested here is sub-fraction of the total abundance of DM, while our bounds do not.\\

\noindent\textit{\textbf{Conclusions and Outlook.---}} We presented existing quantum sensors, which have so far not been used to search for DM, as new DM detectors. We pointed out for the first time that such devices allow a probe of DM through excess quasiparticle generation in single quasiparticle devices, and excess power produced in athermal phonon sensors. We considered DM power deposition in these devices, and their already existing measurements, to constrain two types of DM which potentially exist in the Earth. Firstly, we constrained MeV-scale and higher DM masses from the incoming Galactic halo. Secondly, we set limits on the thermalized DM which is already captured and thermalized within the Earth, for MeV-GeV scale DM.

We identified these new DM sensitivities with three different devices. Single quasiparticle devices provide new constraints already, with promise to provide improved results in future, if lower background noise is achieved. The best limit arises from quasiparticle density measurements, in devices aiming for low quasiparticle backgrounds. The quasiparticle density measurement we used from Ref.~\cite{2013NatCo...4.1913R} may also bring new sensitivities in the future. For example, recently Ref.\,\cite{Mannila_2021} measured an even lower quasiparticle density in superconducting Al. However, we do not use their value because our model of quasiparticle tunneling is not applicable to their experimental setup. This is detailed further in the supplementary material.
We also set new constraints using volume-scaled TES bias power measurements, in which the TES is coupled to a large silicon absorber, as per the SuperCDMS detectors. In future, a larger volume absorber, measured with better systematic controls, would be able to provide stronger sensitivities to thermal DM.

 Interestingly, it is not known what currently produces the quasiparticles measured in these devices~\cite{Vepsalainen:2020trd, Cardani:2020vvp, Wilen:2020lgg, https://doi.org/10.48550/arxiv.2012.06137,Mannila_2021}.  We thus conclude that, for plausible DM parameters, this signal could correspond to a DM signal if seen to remain fixed in time, but caution that this requires proper studies of systematics which at this point are lacking, including an accurate model for relating power injection to quasiparticle generation as we discussed above.

Going forward, our work serves as strong motivation to better understand the systematic uncertainties corresponding to some of these measurements, and motivates further exploration with quantum devices to probe the highly abundant, low velocity, thermalized DM population. Moreover, the encouraging results obtained here will inspire future study to optimize the absorber material for low velocity DM detection. \\

\noindent\textit{\textbf{Acknowledgments.---}} We thank Simon Knapen, Tongyan Lin, Robert McDermott, Britton Plourde, Matt Pyle, and Juri Smirnov for helpful discussions and comments. AD and RKL are supported by the U.S. Department of Energy under Contract DE-AC02-76SF00515. NK is supported by the US. Department of Energy Early Career Research Program (ECRP) under FWP 100872.

\clearpage
\newpage
\maketitle
\onecolumngrid
\begin{center}
\textbf{\large Dark Matter Induced Power in Quantum Devices}

\vspace{0.05in}
{ \it \large Supplemental Material}\\ 
\vspace{0.05in}
{Anirban Das, Noah Kurinsky, and Rebecca K. Leane}
\end{center}
\onecolumngrid
\setcounter{equation}{0}
\setcounter{figure}{0}
\setcounter{section}{0}
\setcounter{table}{0}
\setcounter{page}{1}
\makeatletter
\renewcommand{\theequation}{S\arabic{equation}}
\renewcommand{\thefigure}{S\arabic{figure}}
\renewcommand{\thetable}{S\arabic{table}}

\tableofcontents

\section{Dark Matter Power Deposition: Thin Film or Substrate?}
\begin{figure}[b]
    \centering
    \includegraphics[width=0.47\columnwidth]{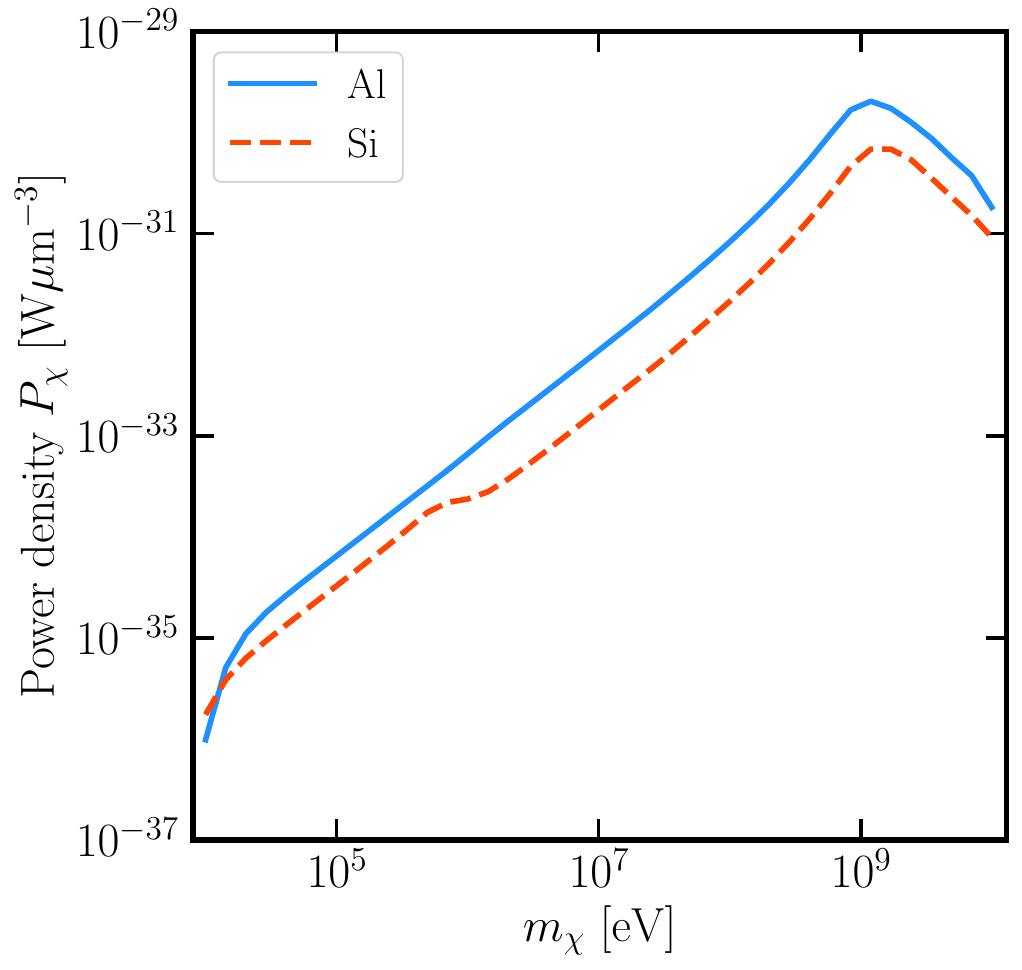}\hspace{4mm}
    \caption{Deposited power density $P_\chi$ as a function of DM mass $\mchi$ for cross section $\sigma_{\chi N}=10^{-34}\cmsq$ through phonon excitation in Al (blue, solid) and Si (red, dashed) crystal, for halo DM.}
    \label{fig:power_mDM}
\end{figure}

The experiments considered in this paper are all small-scale, and often the main absorber volume, i.e. the part of the device that is assumed to absorb energy coming from any external source, is comparable in size with other parts of the experimental setup. Figure~\ref{fig:power_mDM} shows the expected power density in Al and Si from DM scattering as a function of DM mass. We see that the expectations using these two materials differs only by a factor of few; we show the power density for halo DM, though the thermalized DM also only exhibits a factor of a few difference between the two materials. Therefore, a question may arise if one should also take these other components, typically made of different materials than the absorber itself, into account when calculating the power deposit by DM scattering. We address this question in this section.

We consider energy deposition by DM-nuclear scattering through phonon excitation. In case of the transmon qubit (TQ)\,\cite{2013NatCo...4.1913R} and the low noise bolometer (LNB)\,\cite{QCD}, the absorber material is superconducting Al. In the former, the Al film was patterned on a sapphire substrate which leaves a possibility for the phonons, created in the substrate from DM scattering, traveling to the qubit and break Cooper pairs. In this work, we conservatively only consider the volume of the superconductor. Including the substrate volume will only improve our quasiparticle limit. 
In case of the LNB, the Al absorber mesh is separated from the silicon substrate by titanium and gold layers which would stop the phonons from getting to the absorber. The mesh absorber is also a part of the circuit that consists of a resonator made of superconducting niobium, which we also neglect as it acts as a phonon sink\,\cite{QCD}.

We stress that, for direct quasiparticle production, the picture is more subtle than the counting of broken cooper pairs, and refer the reader to Ref.~\cite{bespalov} for a discussion of the dynamics of quasiparticle equilibriation, which leads to the relation shown in Eq.~(\ref{eq:qp_density}). Unlike for semiconductor detectors, the system cannot be considered to be frozen out, and thus any heating will change the effective temperature of the film. In the case that the cooling is limited by the dynamics of quasiparticle recombination, and not the thermalization with the bath, which is the case for the Al devices, the energy injection is not proportional to number of quasiparticles broken by a given power injection rate. Instead, recombination is more efficient at higher generation rates, and thus a sub-linear trend exists between the equilibrium quasiparticle density and quasiparticle breaking power injection.

In contrast, we want to emphasize that the design of the phonon calorimeter used in Refs.\,\cite{Fink_2021,Ren_2021} is substantially different from the other experiments. Here, a much bigger Si substrate block was covered with thin superconducting Al fins. This design was optimized to increase the efficiency of the collection of the ballistic phonons from the Si block into the lower heat capacity Al. Hence, clearly the volume of the Si absorber would dominate the phonon contribution and so we only take that into account. In this case, the Al film mediates down-conversion of phonons in the Si subtrate, and the Al film is thermalized by the lower-Tc TES. The Si substrate is otherwise frozen out for relevant energy scales, and the measurement is sensitive to energy absorbed, rather than quasiparticle density or tunneling rate. The measurement of power injection is therefore more straightforward as compared with the case of measurement via quasiparticle density. 

\section{Details of Phonon Production}\label{sec:details}

\subsection{Formalism}

When the DM energy is low enough that the typical momentum transfer is comparable to the average inverse width of nuclear wavefunction ($q_\mathrm{nuc}\sim \sqrt{2m_A\bar{\omega}}$) in the detector material, collective excitation of the atoms becomes important in the calculation of the response of the detector. For thermalized DM near the surface of the Earth, the typical energy is $E_\mathrm{DM}\simeq 26\meV$ and momentum $q\gtrsim 10\keV$ which is well within the regime where lattice vibrations or phonon excitations due to DM scattering become important (similarly for light DM from the halo). As we will now show, both the energy and momentum play significant role in determining the scattering rate.

The momentum transfer $q$ dictates how the atoms from different lattice sites respond collectively. Following Ref.\,\cite{osti_6632171,Campbell-Deem:2022fqm}, we use the value of $q$ to separate the scattering into two regimes. When $q<\qbz(=2\pi/a)$, the phonon wavelength is greater than the lattice size $a$. In this case, the excitation extends over multiple atoms and the scattering is \emph{coherent}. However, for $q>\qbz$ the phonons have short wavelength and responses from different lattice sites do not interfere. This is the \emph{incoherent} scattering regime.

\begin{figure}[!t]
    \centering
    \includegraphics[width=0.42\columnwidth]{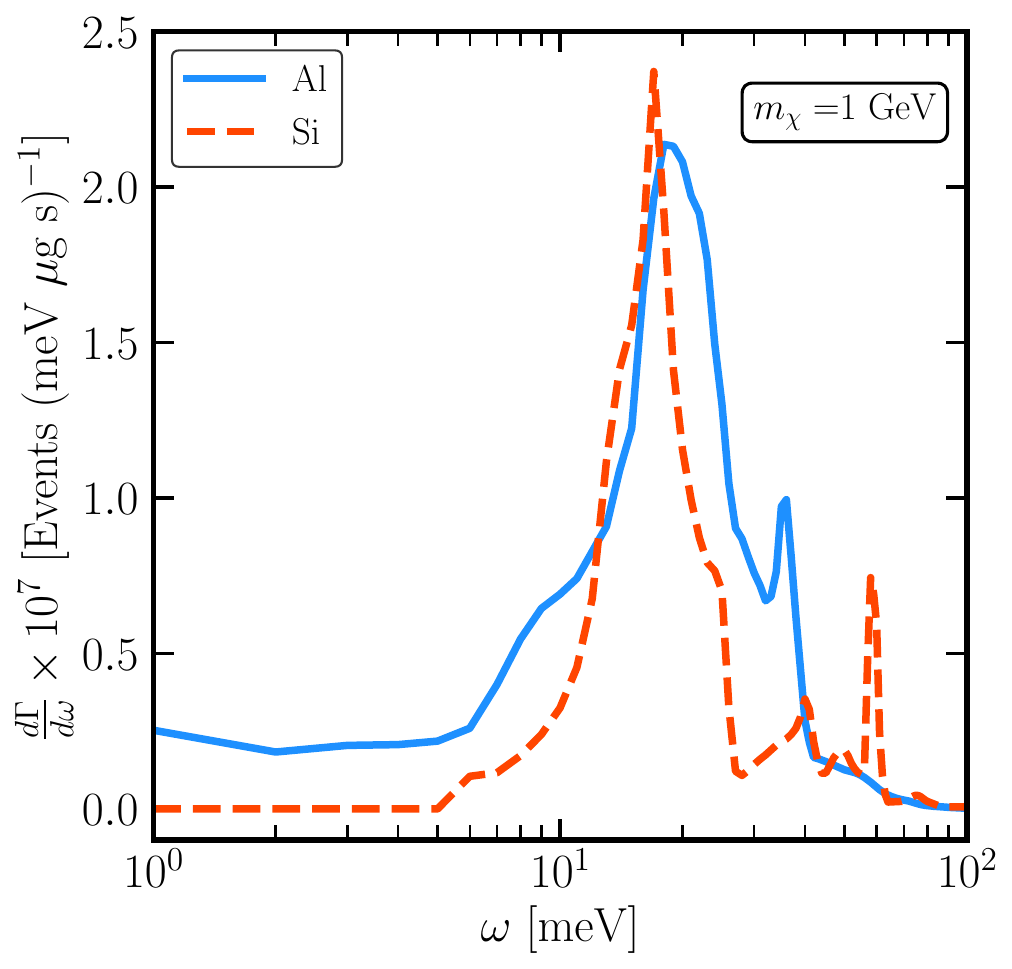}
    \caption{Differential scattering rate for thermalized DM in Al and Si for $\mchi=1\GeV$ and $\sigman=10^{-30}\cmsq$.}
    \label{fig:diff_rates}
\end{figure}

In the most general case, the structure factor $S(q,\omega)$ can be written as a sum over the responses from all lattice sites and atoms of a homogeneous crystal,
\begin{align}\label{eq:s}
    S(\bq,\omega) = \sum_{l,l'}^N \sum_{d,d'}^m \overline{f_{d'}^*f_{d}} C_{l'd'ld}\,.
\end{align}
Here, $l$ is the lattice index, $d$ is the inequivalent atom index within a primitive cell, $N$ is the number of primitive cells in a volume $V$, $m$ is the number of inequivalent atoms in a primitive cell, and $C_{l'd'ld}$ is the time-dependent two-point correlation function for lattice points $l$ and $l'$, atoms $d$ and $d'$ within them,
\begin{align}\label{eq:cld1}
    C_{l'd'ld} \equiv \frac{1}{V} \int_{-\infty}^{+\infty} dt\, \langle e^{-i\bq\cdot\mathbf{u}_{l'd'}(0)} e^{i\bq\cdot\mathbf{u}_{ld}(t)} \rangle e^{-i\omega t}\,,
\end{align}
with $\mathbf{u}_{ld}$ as the displacement vector of the $d$-th atom in $l$-th lattice cell. For incoherent regime, we are not interested in the interferences in the sum in Eq.~(\ref{eq:s}). Hence,
\begin{align}\label{eq:s_inc}
    S(\bq,\omega)\approx \sum_l^N \sum_d^m \left( \overline{f_d^2}-(\overline{f_d})^2\right) C_{ld}\,.
\end{align}
Upon further simplification, Eq.~(\ref{eq:cld1}) takes the form,
\begin{align}\label{eq:cld2}
    C_{l'd'ld} \equiv \frac{1}{V} \int_{-\infty}^{+\infty} dt\, e^{-2W_d(\bq)} e^{\langle \bq\cdot\mathbf{u}_{l'd'}(0) \bq\cdot\mathbf{u}_{l'd'}(t) \rangle} e^{-i\omega t}\,.
\end{align}
After quantizing the lattice displacement vector $\mathbf{u}_{ld}$, the quantity  $\langle \bq\cdot\mathbf{u}_{l'd'}(0) \bq\cdot\mathbf{u}_{l'd'}(t) \rangle$ can be simplified to
\begin{align}
    \langle \bq\cdot\mathbf{u}_{l'd'}(0) \bq\cdot\mathbf{u}_{l'd'}(t) \rangle &\approx \dfrac{q^2}{3}\sum_\nu\sum_\mathbf{k} \frac{|\mathbf{e}_{\nu,\mathbf{k},d}|^2}{2Nm_d\omega_{\nu,\mathbf{k}}} e^{i\omega_{\nu,\mathbf{k}}t}\\
    &= \dfrac{q^2}{2m_d} \int_{-\infty}^{+\infty} d\omega\, \dfrac{D_d(\omega)}{\omega} e^{i\omega t}\,.
\end{align}
Here, $m_d$ is the mass of the atom, $D_d(\omega)$ is the partial phonon DoS of $d$-th atom. This expression can be used to write a general form of $C_{ld}$ for $n$ number of phonon excitation with isotropic assumption,
\begin{align}\label{eq:cld_inc}
    C_{ld} = \dfrac{2\pi}{V} e^{-2W_d(\bq)} \sum_n \dfrac{1}{n!} \left(\dfrac{q^2}{2m_d}\right)^n \left(\prod_{i=1}^n \int d\omega_i \dfrac{D_d(\omega_i)}{\omega_i}\right) \delta\left(\omega-\sum_i\omega_i\right)\,.
\end{align}
with the Debye-Waller function given by
\begin{align}\label{eq:debye-waller}
    W_d(\bq) = \dfrac{q^2}{4m_d} \int d\omega\, \dfrac{D_d(\omega)}{\omega}\,.
\end{align}
Finally, using Eq.~(\ref{eq:cld_inc}) in Eq.~(\ref{eq:s_inc}) yields the structure factor
\begin{align}\label{eq:structure_factor}
    S(\bq,\omega) &\approx \dfrac{2\pi}{V_c} \sum_df_d^2 e^{-2W_d(\bq)} \sum_n \left(\dfrac{q^2}{2m_d}\right)^n 
     \dfrac{1}{n!}\left(\prod_{i=1}^n \int d\omega_i \dfrac{D_d(\omega_i)}{\omega_i}\right) \delta\left(\omega-\sum_i\omega_i\right)\,.
\end{align}
Here $V_c$ is the primitive cell volume and $n$ is the number of phonons excited. The average phonon energy $\bar{\omega}_d$, determined by the DoS of the material, determines the typical number of phonons, $n\sim q^2/(2m_d\bar{\omega}_d)$. In Eq.~(\ref{eq:structure_factor}), we take $f_d=A_d$ as the coupling for spin-independent interactions, where the scattering benefits from nuclear coherence. The differential rate as a function of deposited energy $\omega$ can be written as
\begin{align}\label{eq:rate_spectrum}
    \frac{d\Gamma}{d\omega} = \frac{\pi\sigman\nchi}{\rho_T\mu^2} &\int d^3v\, f_\chi(\bv) \int \frac{d^3q}{(2\pi)^3}\, \fmed^2(q)\, S(\bq,\omega)\, \delta(\omega-\oq)\,.
\end{align}
Figure~\ref{fig:diff_rates} shows example event rates for thermalized DM in both Al and Si, scattering cross sections of $\sigman=10^{-30}\cmsq$. We obtain our event rates by using Eq.~(\ref{eq:rate_spectrum}).

\subsection{The Phonon Density of State and Structure Factor}

\begin{figure*}[!t]
    \centering
    \begin{subfigure}{
    \includegraphics[width=0.45\columnwidth]{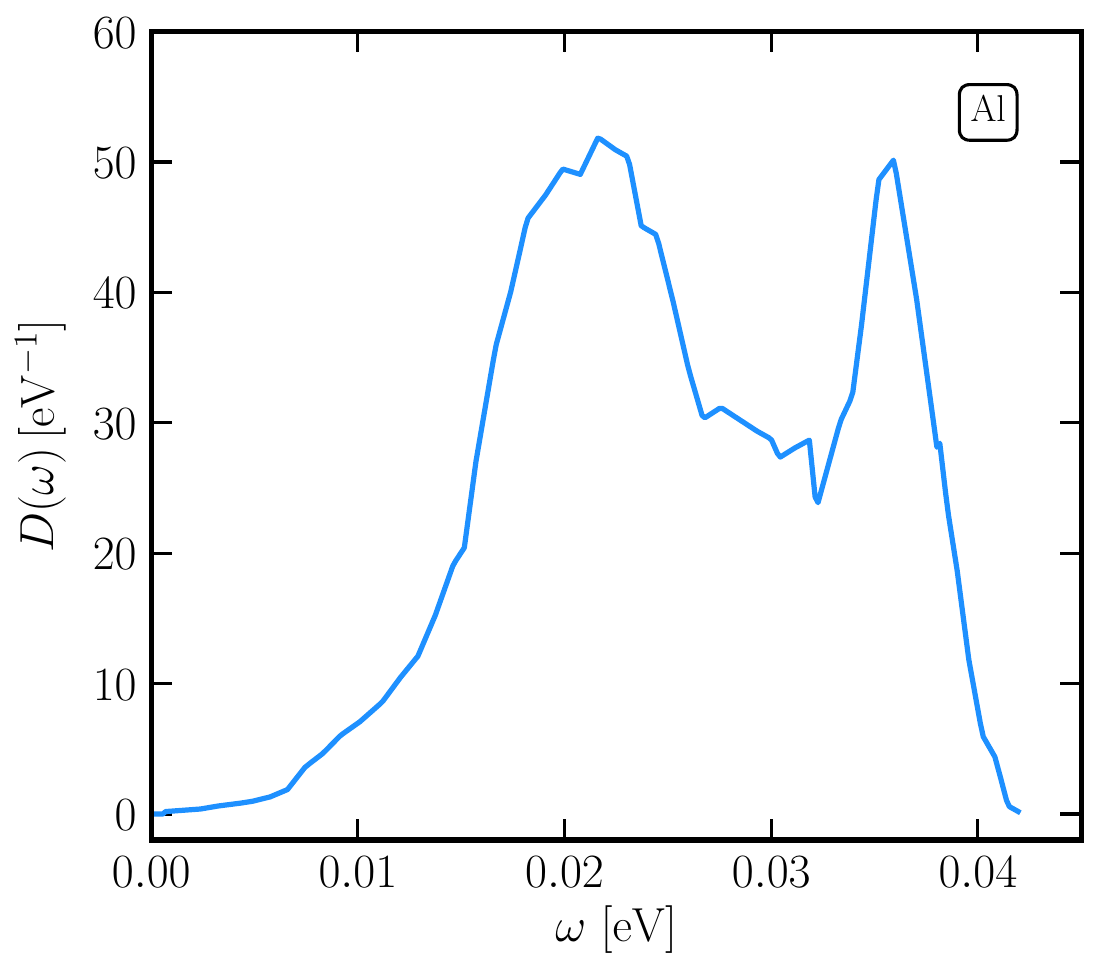}}
    \end{subfigure}
    \begin{subfigure}{
        \includegraphics[width=0.455\columnwidth]{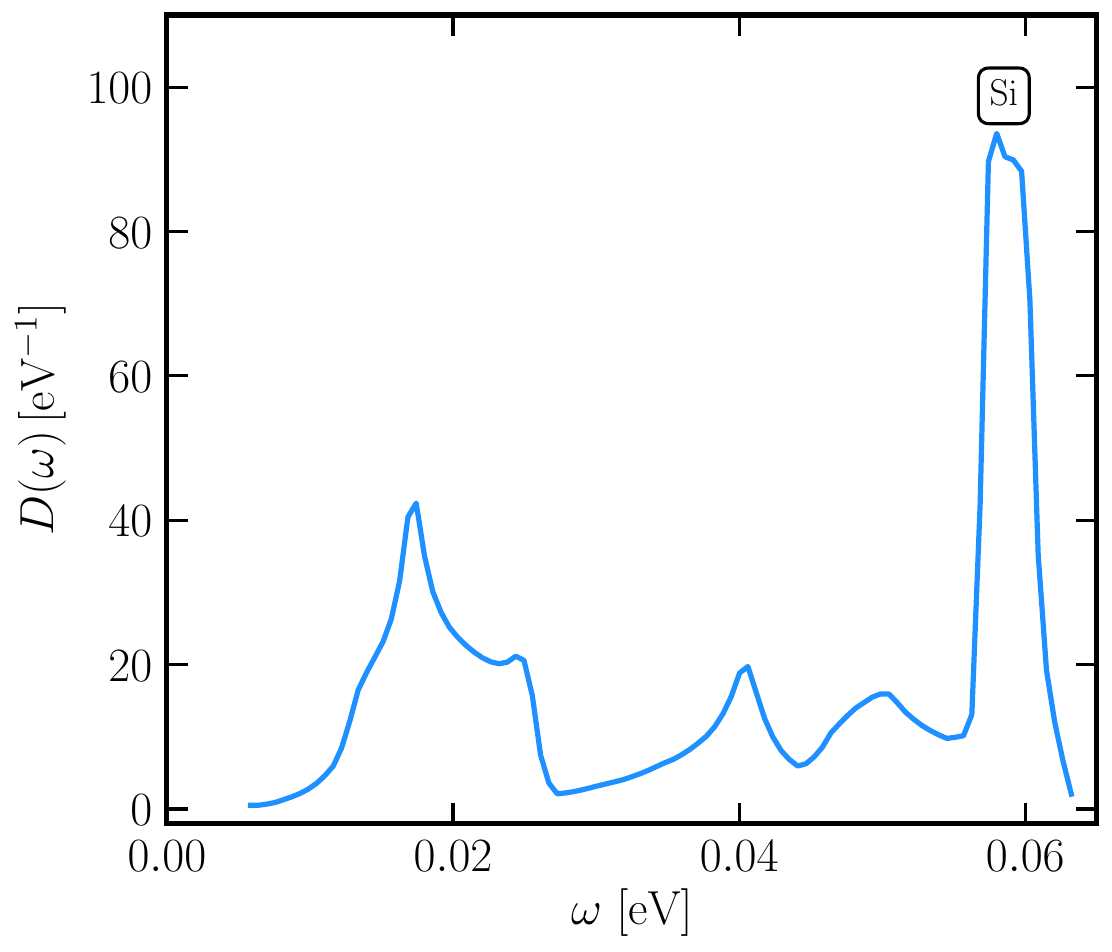}}
    \end{subfigure}
    \caption{The phonon density of state as a function of energy, for aluminum (\textit{left}) and silicon (\textit{right}). For Al, the features are due to the transverse and longitudinal acoustic phonon branches. In Si, the rightmost peak is due the optical phonon branch.}\label{fig:dos}
    \label{fig:my_label}
\end{figure*}

Figure~\ref{fig:dos} shows the phonon DoS $D(\omega)$ for both aluminium (left) and silicon (right), which are the detector materials we have focused on. Al has a face-centered-cubic crystal structure with only one atom in the primitive cell. Hence, it has only an acoustic phonon branch, which is the phonon branch when all atoms oscillate in phase. The two clear peaks in the Al DoS are due to the transverse and longitudinal acoustic branches. Si, on the other hand, has diamond crystal structure with two inequivalent atoms in its primitive cell. Hence, in addition to acoustic branches, it has optical phonon branches, which arise when neighboring atoms oscillate in opposite phase. Compared to acoustic phonon branches, the optical branch adds more phonon states at higher energies~\cite{Kittel2004}, although low-velocity thermalized DM does not exploit this feature. The rightmost peak at $\omega\simeq 0.06\eV$ is due to the optical branches.  The phonon DoS is important for understanding which energies can have resonant energy transfers and result in large scattering rates. For example, the features in the differential scattering rates in Fig.~\ref{fig:diff_rates} can be associated with the corresponding features in the DoS shown above. The $\sim 1/\bar{\omega}_d$ dependence of $S(q,\omega)$ in Eq.~(\ref{eq:structure_factor}) suppresses the rate at higher energy.

\begin{figure*}[!t]
    \centering
    \begin{subfigure}{
        \includegraphics[width=0.45\columnwidth]{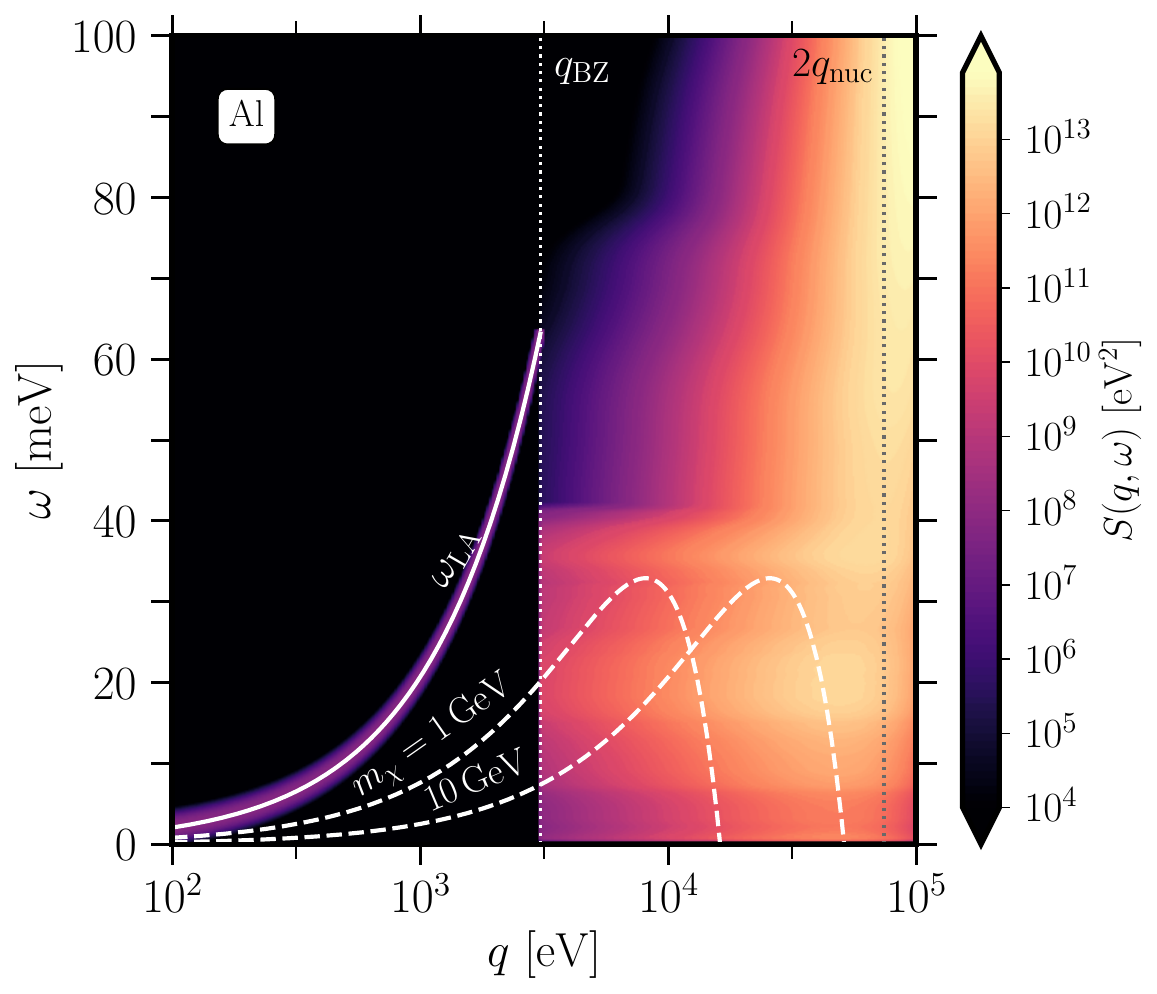}}
    \end{subfigure}
    \begin{subfigure}{
        \includegraphics[width=0.45\columnwidth]{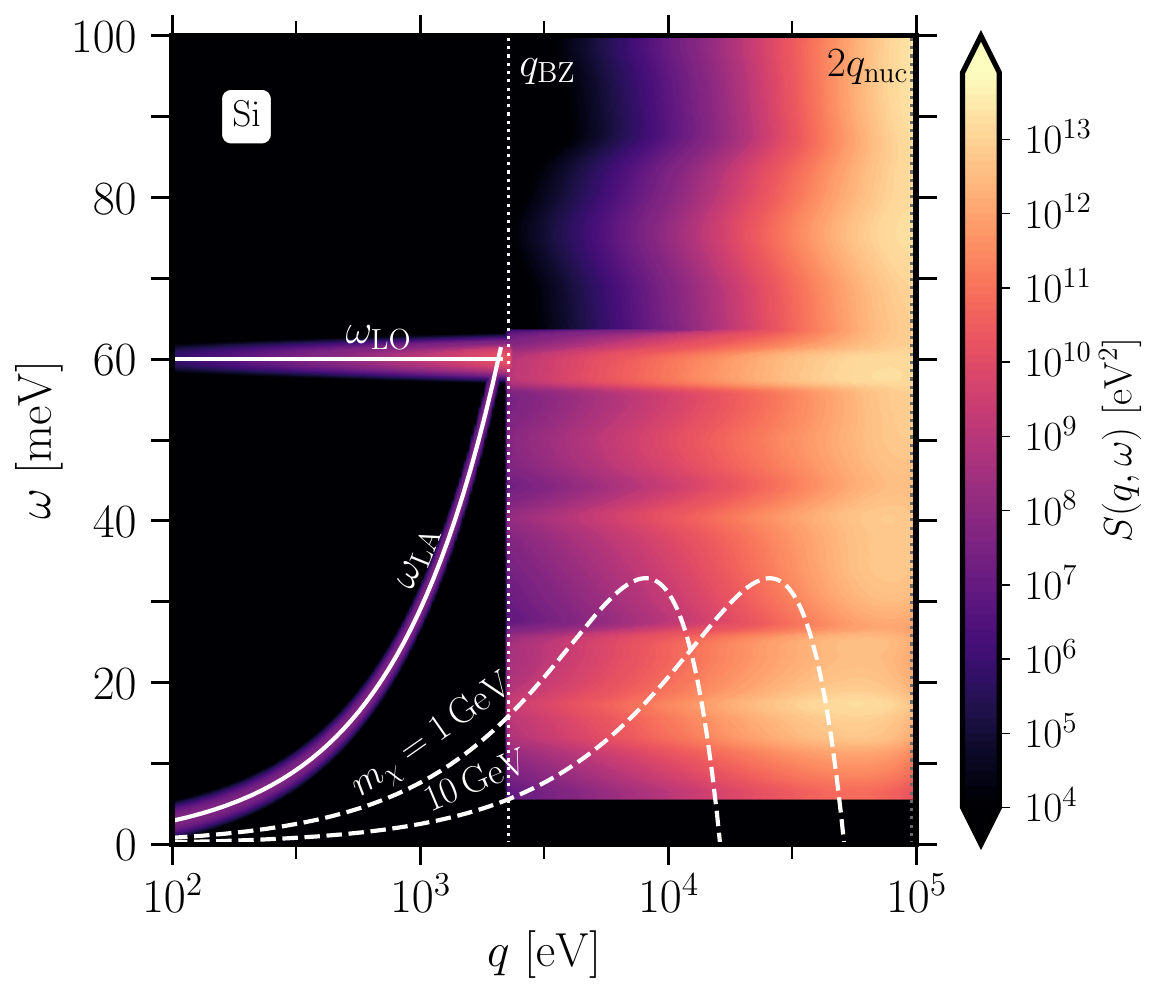}}
    \end{subfigure}
    \caption{The incoherent structure factor $S(q,\omega)$ as a function of momentum $q$ and phonon energy $\omega$ for aluminum (left) and silicon (right). Below the inverse lattice constant $\qbz$ the scattering is incoherent. The longitudinal acoustic and optical branches, labeled as $\omega_{\rm LA}$ and $\omega_{\rm LO}$ respectively, are shown in solid white lines. We broadened the structure factor along these lines with a $0.5\meV$ due to finite phonon decay width. Above $\qbz$ in the incoherent regime, the bright bands are due to the varying number of phonon excitations. Above the inverse length of the nuclear wavefunction $q_{\rm nuc}$, the scattering can be approximated through nuclear recoils rather than phonon excitations. The thermalized DM phase spaces for $\mchi=1$ and $10\GeV$ are marked with dashed white lines.}\label{fig:phase_space}
\end{figure*}

Figure~\ref{fig:phase_space} shows the incoherence structure factor for aluminium (left) and silicon (right), as well as the DM phase space for thermalized DM with $\mchi=1$ and $10\GeV$. Below $\qbz$, only narrow regions along the $\omega_\mathrm{LA}$ and $\omega_\mathrm{LO}$ lines can be excited. We choose a finite width of $0.5\meV$ along the lines which approximates realistic phonon decay lifetimes. At low momentum, the acoustic branch has a dispersion $\omega\sim q$, whereas the optical branch dispersion is assumed to be constant in $q$ for simplicity. Above $\qbz$, the structure factor broadens due to multiple phonon excitations. 

In Fig.~\ref{fig:phase_space}, the DM phase spaces are marked with white dashed lines (see Eq.~(\ref{eq:DM_phase_space}) for the definition). These shapes vary greatly from galactic halo DM phase spaces, as thermalized DM does not increase its maximum energy with mass as it is thermalized with temperature $T_\chi\simeq 300\K$. To compare coherent and incoherent scattering regimes, we note that the lattice constants of Al and Si are $4.05\,$\AA{} and $5.43\,$\AA, respectively. They correspond to Brillouin zone boundaries $\qbz=3.06\keV$ and $2.3\keV$ in the Fourier space. Therefore, even though thermalized DM has lower kinetic energy than the galactic halo DM, these momenta fall below the typical momenta of such DM of mass between 1 and $10\GeV$. Moreover, a large part of them lie above $\qbz$ in both cases. Therefore, the incoherent scattering dominates the scattering rate.

\subsection{Number of Phonons Excited}

It is also interesting to ask the average number of phonon excited per single DM scattering. This can be estimated by $n\sim q^2/(2m_A\bar{\omega})$. We find $\bar{\omega}=25.5$ and $40.8\meV$ for Al and Si, respectively. For $\mchi=1\GeV$ with a maximum momentum transfer of $q=20\keV$, this yields $n = 0.29$ and $0.18$, respectively. For comparison, halo DM of the same mass can have maximum momentum $q=0.73\MeV$ corresponding to $n\gg 1$. This is because thermalized DM has less kinetic energy than its halo counterpart. Finally, we note that because of the lower energy, the overlap between the DM phase space and $S(q,\omega)$ is not optimal for either of Al and Si. A material with more states at low energy $\omega\lesssim 30\meV$ could yield higher energy transfer to the detector from thermalized DM. We leave further investigation in this direction to future work.

\section{Regimes of Validity: Phonon Excitation or Nuclear Recoil?}

Finally, we would like to emphasize that the relevant degrees of freedom in a detector material depends on the typical momentum and energy of the scattering DM particles. The critical momentum scale in this case is the average inverse width of nuclear wavefunction $q_\mathrm{nuc}\sim \sqrt{2m_A\bar{\omega}}$~\cite{Trickle:2019nya,Campbell-Deem:2022fqm,Kahn:2021ttr}. Note that $q_\mathrm{nuc}=37.1$ and $47.8\keV$ for Al and Si respectively. If the typical DM momentum $q_\mathrm{DM} \lesssim q_\mathrm{nuc}$, then the energy deposition occurs mainly by phonon excitation, and the atoms are not free particles. On the other hand, only when $q_\mathrm{DM} > 2q_\mathrm{nuc}$, single nuclear recoil becomes an appropriate description of the scattering process. Thermalized DM with $T_\chi \simeq 300\K$ can have a maximum momentum transfer
\begin{align}
    q_\mathrm{max} = 6.37\keV~\left(\dfrac{\mchi}{1\GeV}\right)^\frac{1}{2}\,.
\end{align}
Clearly, this is smaller than $q_\mathrm{nuc}$ for all DM masses below $\mchi \lesssim 35\GeV$. Therefore, phonon excitation is the appropriate treatment for our parameter space.

\section{Power Measurement in Low Quasiparticle Background Devices}
\label{app:comp}

The main results involving interpretation of quasiparticle tunneling rates rely on relating a power injection to the equilibrium quasiparticle density. Here we consider two cases: a discrete event model (tunneling rate tracks event rate) and a rate-balance model (generation rate due to power injection tracks equilibrium density). The relevant model affects the computed power constraint, and both need to be considered to understand the range of potential DM models which can be probed.

Suppose first that we measure discrete events which are Poisson-distributed both in energy and time with a mean rate $\Gamma_0$ and mean energy $\lambda_E$. We have a time-averaged power equal simply to $\Gamma_0\lambda_E$, and convert this to a volumetric time-averaged power as $\Gamma_0\lambda_E/V$. According to Ref.~\cite{Mannila_2021}, $\Gamma_0 = 1.5 \,\mathrm{Hz}$ and $ \lambda_{qp} = 0.9$ as the number of Cooper pairs broken per event in a volume of $0.035\mum^{3}$. The Cooper pair breaking energy is $220\mueV$, which allows us to calculate the total energy dissipated as
$\lambda_E = \lambda_{qp} \times 220\mueV = 198\mueV = 3.5\times 10^{-23}\,{\rm J}$. We thus find a mean power injection density of $10^{-21}\Wmum$. However, a major caveat to this point is that the typical energy of those events is at a much lower energy scale than the expected DM events if we assume they can count all excess quasiparticles produced in the island - this uncertainty makes this power less clearly a constraint, and more likely an intermediate estimate of a range of potential power injections.

On the other hand, the power density calculated using the mean field theory is only $6\times 10^{-25}\Wmum$ for the same number density. We can understand the difference between these numbers as follows. The quasiparticle number density evolution follows the Boltzmann equation,
\begin{equation}
    \dfrac{d\nqp}{dt} = -\Gamma_R - \Gamma_T + A \approx -\Bar{\Gamma}\nqp^2 - \Bar{\Gamma}_T\nqp + A\,,
\end{equation}
where $\Gamma_R, \Gamma_T, A$ are the recombination, trapping, and generation rates, respectively. For a steady state pair-breaking power density $P$, we can write $A=P/(2\Delta)$ where $\Delta$ is the gap energy. Therefore, in equilibrium
\begin{equation}
P/(2\Delta) = \Bar{\Gamma}\nqp^2 + \Bar{\Gamma}_T\nqp \,.
\end{equation}
No QP trapping is assumed in the mean field calculation with $\Bar{\Gamma}_T=0$, leading to the relation $\nqp=\sqrt{A/\Bar{\Gamma}}\propto \sqrt{P}$. However, we find a different scaling when the trapping is non-negligible, $\nqp\approx A/\Bar{\Gamma}_T\propto P$. 

Measurement of the tunneling rate as a function of DC power injection allows for quick determination of the relevant regime for quasiparticle dynamics, as is done in Ref.~\cite{QCD}. This is not done in Ref.~\cite{Mannila_2021}, and thus the mean field calculation of the QP density is not readily applicable to it. We thus do not use their data in this letter.

\bibliography{DM_noise}
\bibliographystyle{apsrev4-2}

\end{document}